\documentclass[aps,twocolumn,nofootinbib]{revtex4}

\usepackage{graphicx}
\usepackage{dcolumn}
\usepackage{bm}

\usepackage{amssymb,amsfonts}
\usepackage{epsfig}
\usepackage{epstopdf}
\usepackage[all]{xy}
\usepackage{amsthm}
\usepackage{dcolumn}
\usepackage{hyperref}
\usepackage{url}
\usepackage{textcomp}
\usepackage{dsfont}

\newcommand{\be}{\begin{equation}}
\newcommand{\ee}{\end{equation}}
\newcommand{\bea}{\begin{eqnarray}}
\newcommand{\nn}{\nonumber}
\newcommand{\eea}{\end{eqnarray}}

\def\inbar{\,\vrule height1.5ex width.4pt depth0pt}
\def\IR{\relax{\rm I\kern-.18em R}}
\def\IC{\relax\hbox{$\inbar\kern-.3em{\rm C}$}}

\begin{document}

\title{Casimir energy-momentum tensor for a quantized bulk scalar field\\ in the geometry of two curved branes on Friedmann-Robertson-Walker background}

\author{Hamed Pejhan\footnote{h.pejhan@piau.ac.ir}}
\affiliation{Department of Physics, Science and Research Branch, Islamic Azad University, Tehran, Iran}

\author{Surena Rahbardehghan\footnote{s.rahbardehghan@iauctb.ac.ir}}
\affiliation{Department of Physics, Islamic Azad University, Central Tehran Branch, Tehran, Iran}

\begin{abstract}
In a previous work \cite{Suren750}, we considered a simple brane-world model; a single $4$-dimensional brane embedded in a $5$-dimensional de Sitter (dS) space-time. Then, by including a conformally coupled scalar field in the bulk, we studied the induced Casimir energy-momentum tensor. Technically, the Krein-Gupta-Bleuler (KGB) quantization scheme as a covariant and renormalizable quantum field theory in dS space was used to perform the calculations. In the present paper, we generalize this study to a less idealized, but physically motivated, scenario, namely we consider Friedmann-Robertson-Walker (FRW) space-time which behaves asymptotically as a dS space-time. More precisely, we evaluate Casimir energy-momentum tensor for a system with two $D$-dimensional curved branes on background of $D+1$-dimensional FRW space-time with negative spatial curvature and a conformally coupled bulk scalar field that satisfies Dirichlet boundary condition on the branes.
\end{abstract}
\maketitle

\section{Introduction}
\label{sec:intro}
In the semi-classical approach to quantum gravity, mean values of the energy-momentum tensor act as the source of gravity in the Einstein equation. Therefore, in modelling a self-consistent dynamics involving the gravitational field, such as a self-consistent formulation of the brane-world dynamics, it plays an essential role.

As a matter of fact, Brane-world models\footnote{In the original brane-world scenarios such as the so-called RS models, standard model particles and fields are localized on a D-dimensional hyper-surface called the "brane" embedded in a higher-dimensional space-time called the "bulk", where gravity is the only field which has access to the extra dimensions.} and their generalizations (to include matter on the brane, scalar fields in the bulk, etc.) have always been an interesting option within extra-dimensional theories due to their attractive features and applications (for a review of brane-world scenarios refer to \cite{Langlois163,Maartens13,Shiromizu62,Brax67,Goldberger107505}). This point of view of our universe may bring about interesting mechanisms to resolve such well-known problems as the cosmological constant and the hierarchy problems \cite{Wasserman86,Randall3370}. These scenarios can also lead to inflationary or late time acceleration models of the universe \cite{Ma03,Wong241,Brodie12}. In addition, the brane-world approach, as manifestation of the holographic principle, is a nice model to study (A)dS/CFT correspondence \cite{Duff85,Hebecker608,Nojiri494,Shiromizu64,Molina82,Nojiri0112,Petkou0202,Strominger0110,Myung531}. Therefore, the investigation of quantum effects in brane-world models is of significant concern, both in particle physics and in cosmology, which must be considered in a self-consistent formulation of the brane-world dynamics.

An intrinsic feature of these scenarios is the presence of boundaries and propagating fields in the bulk, which will naturally give Casimir-type contributions to the mean values of physical observables (to review the Casimir effect refer to \cite{Elizalde}). In particular, vacuum forces arise acting on the branes, which depending on the type of fields and imposed boundary conditions, can either stabilize or destabilize the brane-world (it is directly related to the hierarchy problem). Moreover, the Casimir energy allocates a contribution to both the bulk and the brane cosmological constants, and therefore, one expects that it can be helpful in the resolution of the cosmological constant problem.

In such semi-classical theories of quantum gravity in order to evaluate mean values of the energy-momentum tensor, however, it seems that a consistent formulation with the minimal requirements of an acceptable quantization procedure of the (linear) quantum gravity on the considered background (bulk) should be in order. Amongst other physically motivated space-times (backgrounds), de Sitter space, as the maximally symmetric solution of Einstein equation (with a positive cosmological constant), becomes important at large-scale universe. It is likely that universe evolves into the future (asymptotically) de Sitter phase; the recent astronomical observations of supernova and cosmic microwave background \cite{Riess} indicate that our universe is accelerating and can be well approximated by a world with a positive cosmological constant. The quantum field theory on dS background is also of considerable interest; it is maximally symmetric curved space-time and provides the opportunity of controlling the transition to the flat space-time by the so-called contraction procedure (see \cite{Garidiprd} and references herein). Nevertheless, even on this very simple curved background, one encounters difficulties in defining quantum fields, including the free fields.

A famous example is the so-called dS massless minimally coupled scalar field. [It plays a central role in the inflationary scenario \cite{Linde2} and the linear quantum gravity in dS space \cite{Ford1601}.] According to Allen's theorem \cite{Allen,AllenFolacci}, for this field in dS space, due to the advent of infrared (IR) divergence, there is no Hilbertian dS-invariant Fock vacuum state. Therefore, no covariant canonical quantization, through a Fock construction based upon a one-particle sector of the Hilbert space structure, is possible. On the other hand, as is well-known, the graviton propagator on dS background behaves in a manner similar to that for the minimally coupled scalar field \cite{Ford1601}. This similarity results in the same difficulty for the graviton field on dS background (the dS linear gravity). Indeed, respecting the standard QFT, it is generally accepted that, for these fields, the phenomenon of dS-breaking is universal and the corresponding propagators suffer from IR divergences \cite{Allen,AllenFolacci,Allen813,Floratos373,Antoniadis1037,Mora53}.

Recently, however, thanks to a rigorous group theoretical approach combined with a suitable adaptation (Krein spaces) of the Wightman-G\"{a}rding axiomatic for massless fields (Gupta-Bleuler scheme) \cite{Wightman}, the so-called Krein-Gupta-Bleuler (an indefinite metric field) quantization method, a causal and fully covariant quantization of the minimally coupled massless scalar field and consequently linearized gravitons on dS background has been presented (it is free of any infrared divergence) \cite{de Bievre6230,Gazeau1415,Garidi,Bertola,Hamed} (in this regard, see also \cite{Dehghani064028,Pejhan2052,Rahbardehghan}). Indeed, contrary to the usual approach to the dS linear gravity, it is proved that the KGB construction of the graviton propagator, at least in the absence of graviton-graviton interaction, does not suffer from the pathological large distance behavior (the dS-covariance is indeed preserved). For detailed discussions see Ref. \cite{Hamed}.

Motivated by the above statements, in a previous work \cite{Suren750}, based on the Krein method, we studied the bulk Casimir effect for a conformally coupled scalar field when the bulk represents 5-dimensional de Sitter space-time with one 4-dimensional dS brane. As already mentioned, using the KGB quantization, in general sense, is the necessity of being consistent with the covariance requirements of the minimally coupled field and the linear gravity on dS background. Moreover, utilizing the KGB formalism also brings an automatic and covariant renormalization of the mean values of the energy-momentum tensor (Wald axioms are well preserved \cite{Gazeau1415}), which can be a remarkable advantage for the theory, especially when dealing with quantizing fields on curved backgrounds.

Pursuing this path, in the present work, in order to extend the theory, we investigate a more realistic model in the geometry of two $D$-dimensional curved branes with Dirichlet boundary condition in the less symmetric $D+1$-dimensional Friedmann-Robertson-Walker bulk, by calculating expectation values of the energy-momentum tensor for a conformally coupled bulk scalar field through the Krein quantization formalism. Friedmann-Robertson-Walker space-time is actually a suitable choice for our purpose to investigate. It describes a homogeneous and isotropic space-time in the context of General Relativity which our universe, over a very large scale, is similar to, and hence it can be considered as a standard model of modern cosmology. Moreover, it asymptotically tends to de Sitter space-time and can also lead to inflationary or late time acceleration models of the Universe.

The layout of the paper is as follows. In section II, we study the free field energy-momentum tensor for a conformally coupled massless scalar field in $D+1$-dimensional FRW space-time with negative spatial curvature. We show that the KGB construction interestingly provides an automatic and covariant renormalization of the expectation values of the energy-momentum tensor. Then in section III, by considering two $D$-dimensional curved branes which are actually the conformal images of two infinite parallel plates moving with constant proper acceleration in Rindler stace-time and admitting Dirichlet boundary condition on their surfaces, the induced Casimir energy-momentum tensor for a quantized conformally coupled bulk scalar field is evaluated. Technically, to fulfil this purpose and to make the maximum use of the flat space-time calculations, we present the FRW line element in the form conformally related to the Rindler metric. Finally, in section IV, a brief summery and discussion is given.

\section{Covariant renormalization of the energy-momentum tensor through the KGB quantization}
In this section, we apply the Krein quantization method to a conformally coupled massless scalar field on background of less symmetric $(D+1)$-dimensional FRW space-time with negative spatial curvature in order to investigate mean values of the energy-momentum tensor.

In this regard, by considering the hyperspherical coordinates $(r,\theta_1=\theta,\theta_2,...,\theta_{D-1})$, the corresponding line element can be written as follows
\begin{eqnarray}\label{metric} ds^2= g_{ik} dx^i dx^k = a^2(\eta)(d\eta^2- \gamma^2dr^2 -r^2d\Omega^2_{D-1}),  \end{eqnarray}
in which $\gamma=1/\sqrt{1+r^2}$ and $d\Omega^2_{D-1}=d\theta^2 + \sum_{k=2}^{D-1}(\prod_{l=1}^{k-1}\sin^2 \theta_l)d\theta^2_k$ is the line element on the $(D - 1)$-dimensional unit sphere in Euclidean space. With regard to the line element (\ref{metric}), for the conformally coupled scalar field, the field equation would be
\begin{equation} \label{Eq} (\nabla_l\nabla^l + \frac{D-1}{4D} R)\phi(x)=0, \end{equation}
$\nabla_l$ and $R$ are, respectively, the covariant derivative and the Ricci scalar for the corresponding metric.

For each regular solution $\phi(x)$ of Eq. (\ref{Eq}), we have
\begin{equation} \phi(x)=\int_\Sigma \widetilde{G}(x,x') {\mathop{\partial_i}\limits^\leftrightarrow }\phi(x') d\Sigma^i (x'),\label{Cauchy} \end{equation}
in which $\Sigma$ stands for the Cauchy space-like surfaces, and the function $\widetilde{G}$˜ refers to the commutator of the field. In consistency with (\ref{Cauchy}), for the solutions space, the inner product is defined by
\begin{equation}\label{inner}(\phi_1,\phi_2)=-i\int_\Sigma\phi_1\,{\mathop{\partial_i}\limits^\leftrightarrow}\,\phi^*_2d\Sigma^i.\end{equation}
With respect to the product (\ref{inner}), there exists a complete orthonormal set of mode solutions of Eq. (\ref{Eq}), i.e.
\begin{eqnarray} &(\phi_\alpha,\phi_{\alpha'})=\delta _{\alpha \alpha'}, \;\; (\phi_\alpha ^*,\phi_{\alpha'}^*)= -\delta _{\alpha \alpha'},& \nn\\
&(\phi_\alpha,\phi_{\alpha'}^*)= 0,& \label{eq:ortho} \end{eqnarray}
the set of $\{\phi_\alpha ,\phi_\alpha ^*\}$ are, respectively, positive and negative norm states. 

In the context of the Krein quantization method, the vacuum state and associated Fock space are constructed by expanding the field operator $\varphi$ in terms of $\{\phi_\alpha,\phi^\ast_\alpha\}$ as follows \cite{Garidi}
\begin{eqnarray} \label{fo-frw} \varphi =\frac{1}{\sqrt{2}}(\varphi_+ + \varphi_-)\hspace{2cm}\nn\\
=\frac{1}{\sqrt{2}} \Big( \sum_{\alpha \in {\cal{K}}} (a_\alpha \phi_\alpha + a^\dagger_\alpha \phi^*_\alpha) + \sum_{\alpha \in {\cal{K}}} (b^\dagger_\alpha \phi_\alpha + b_\alpha \phi^*_\alpha) \Big),
\end{eqnarray}
where ${\cal{K}}$ is a set of indices indexing the positive norm modes, and $\alpha , \alpha' \in {\cal{K}}$. The subscripts $"+"$ and $"-"$ are, respectively, referred to physical and non-physical parts of the field operator. $a_\alpha |0\rangle=0$ and $b_\alpha |0\rangle=0$ determine the Fock vacuum state $|0\rangle$, while $a_\alpha ^\dag |0\rangle=|1_\alpha \rangle$ and $b_\alpha ^\dag |0\rangle= |\bar 1_\alpha \rangle$ are the physical and non-physical one-particle states. Note that, $[a_\alpha, a^\dagger_{\alpha'}] = \delta_{\alpha \alpha'},\; [b_\alpha, b^\dagger_{\alpha'}] = -\delta_{\alpha \alpha'}$ and the other commutation relations are zero.

The Krein-Fock vacuum is normalizable and unique (independent of Bogolioubov transformations) \cite{Gazeau1415}. However, this is no concern, since in this quantization method not only the vacuum but also the field itself is different. The point is indeed laid in the concept of defining observables in the Gupta-Bleuler construction in the sense that the set of states is different from the set of physical states; the observables are defined on the total space, while the average values of the observables are calculated on the sub-space of physical states.

Furthermore, it must be expressed that the invariance of the Fock vacuum does not infer that Bogolioubov transformations, which are merely a simple modification of the set of physical states, are no longer valid in this formalism. As a matter of fact, the physical states space is affected by both the observer and the considered space-time; an accelerated observer in Minkowski space has different physical states from those an inertial observer does (Unruh effect), while for both cases, the same field representation can be utilized \cite{Gazeau1415}. In summary, in the context of the KGB construction, \emph{"instead of possessing a multiplicity of vacua, we have several possibilities for the space of physical states and only one field and one vacuum which do not depend on Bogolioubov transformations. More accurately, the usual ambiguity about vacua is not suppressed but displaced"} \cite{Garidi}.

Now, let us focus on the energy-momentum tensor $T_{ij}$. As already pointed out, in this context as usual in a Gupta-Bleuler formalism, the expectation values of the energy-momentum tensor, as an observable, will be evaluated only with physical states $$|\vec{\alpha}\rangle\equiv|{\alpha}_1^{n_1}...{\alpha}_q^{n_q}\rangle = \frac{1}{\sqrt{n_1!...n_q!}}(a_{\alpha_1}^\dag)^{n_1}...(a_{\alpha_q}^\dag)^{n_q}|0\rangle.$$
In this regard, the starting point generally is
\begin{eqnarray}
\langle \vec{\alpha}|\partial_i\varphi(x)\partial_j\varphi(x)|\vec{\alpha}\rangle = \sum_{\alpha \in {\cal{K}}} \partial_i\phi_\alpha (x)\partial_j\phi_\alpha ^\ast(x)\hspace{1.5cm}\nn\\
- \partial_i\phi_\alpha ^\ast(x)\partial_j\phi_\alpha (x) + 2 \sum_{p = 1}^l n_p \Re \Big(\partial_i\phi_{\alpha_p}^\ast(x)\partial_j\phi_{\alpha_p}(x)\Big).\;\;
\end{eqnarray}
Analogous to the usual QFT, the first term on the right is responsible for appearing divergences in the theory. However, the unusual presence of the second term with the minus sign (corresponding to the terms of the field containing $b_\alpha$ and $b_\alpha ^\dag$) automatically removes divergences and hence there is an automatic renormalization of the $T_{ij}$'s (no infinite term appears). So, we obtain
\begin{equation}
\langle \vec{\alpha}|\partial_i\varphi(x)\partial_i\varphi(x)|\vec{\alpha}\rangle = 2 \sum_{p = 1}^l n_p \partial_i\phi_{\alpha_p}^\ast(x)\partial_i\phi_{\alpha_p}(x).
\end{equation}
Accordingly, the positivity of the energy for any physical state $|\vec{\alpha}\rangle$; $\langle \vec{\alpha}|T_{00}| \vec{\alpha}\rangle\geq0$ ($ = 0 \Leftrightarrow |\vec{\alpha}\rangle=|0\rangle$), as a reasonable physical interpretation of the method, is guaranteed.

This automatic renormalizing process remarkably fulfills the so-called Wald axioms \cite{Gazeau1415}:
\begin{itemize}
\item{The causality and covariance are assured because the constructed field is causal and convarinat.}
\item{The above computations provide the formal results for physical states.}
\item{The foundation of the above computations is ($[b_\alpha,b_\alpha ^\dag]=-1$)
$$a_\alpha a_\alpha ^\dag+a_\alpha ^\dag a_\alpha +b_\alpha b_\alpha ^\dag+b_\alpha ^\dag b_\alpha = 2a_\alpha ^\dag a_\alpha +2b_\alpha ^\dag b_\alpha.$$
So, by applying the method to the physical states (on which $b_\alpha$ vanishes), it can be seen that the procedure is equivalent to reordering.}
\end{itemize}

\textit{Remarks on the renormalization:} In the context of the usual QFT through a Hilbertian-Fock construction, any instruction for renormalizing the energy-momentum tensor, which is consistent with Wald axioms, must yield precisely the trace, modulo the trace of a conserved local curvature term \cite{Wald}. According to the above statements, however, all components of the $\langle T_{ij}\rangle$ vanish in the Krein-Fock vacuum, and hence, the so-called conformal anomaly\footnote{Breaking the conformal invariance when quantum corrections are included.} disappears from the trace of the energy-momentum tensor. Of course, with respect to the fact that by performing the computations through the Krein formalism, covariance and conformally covariance of the quantum field are preserved in a rather strong sense,\footnote{There is only one field and one vacuum which do not depend on Bogolioubov transformations (for a detailed discussion, see \cite{de Bievre6230,Gazeau1415,Garidi}).} therefore it is not surprising that the trace anomaly, which can appear only through the conformal anomaly, vanishes. Once again, we must emphasize that although in the Krein context the trace anomaly does not appear, Wald axioms are well preserved \cite{Gazeau1415}.

Here, it is worth mentioning that in Ref. \cite{HawkingRadiation} by studying black hole radiation, it has been shown that the aspects of the black hole thermodynamics have been emerged in the context of the Krein quantization, even possessing this particular property that the VEV of the energy-momentum tensor is zero.

\section{Evaluating the Casimir energy-momentum tensor}
In this section, the field is supposed to satisfy Dirichlet boundary condition on the surface of two boundaries which are chosen to be two infinite parallel plates moving by uniform proper acceleration in the Rindler right wedge (this will be discussed in a few lines later). These boundaries are characterized in FRW space-time by the equations
\begin{eqnarray}\label{bc} \sqrt{1+r^2} - r\cos\theta = c_s,\;\;s=1,2 \end{eqnarray}
in which $c_{s=1,2}$ are positive constants, so that $c_1>c_2$. Respecting this scenario, the mean values of the energy-momentum tensor in what follows are investigated.

Before coming back to this point and in order to make the maximum use of the flat space computations, we initially perform calculations in Rindler space-time, so let us present the line element (\ref{metric}) in the form conformally related to the Rindler metric. In this regard, we use the following standard coordinate transformation
\begin{eqnarray}\label{ct} x^i = (\eta,r,\theta_1, ... , \theta_{D-1})\rightarrow x'^i = (\eta,x'), \end{eqnarray}
with $x'=(x'^1,x'^2, ... , x'^D)$;
\begin{eqnarray}\label{x'} x'^1 &\equiv &\xi = \Omega\xi_0, \nn\\
x'^2 &=& \Omega\xi_0 r\cos\theta_2\sin\theta, \nn\\
&.&\nn\\
&.&\nn\\
&.&\nn\\
x'^{D-1} &=& \Omega\xi_0 r \cos\theta_{D-1} \prod_{l=1}^{D-2}\sin\theta_l,\nn\\
x'^{D} &=& \Omega\xi_0 r \prod_{l=1}^{D-1}\sin\theta_l. \end{eqnarray}
Note that, in the above identities $\Omega= \gamma/(1-r\gamma\cos\theta)$ and $\xi_0$ is a constant of the length dimension. The line element (\ref{metric}), now, can be expressed in the $x'^i$ coordinates as follows
\begin{eqnarray}\label{conformally} ds^2 = g'_{ij} dx'^i dx'^j = a^2(\eta)\xi^{-2}\bar{g}_{ij} dx'^i dx'^j, \end{eqnarray}
which is manifestly conformally related to the Rindler space metric, $\bar{g}_{ij} = \mbox{diag}(\xi^2,-1, ... ,-1)$.\footnote{From now on, the 'bar' sign indicates that we are working in Rindler space-time.} The equations (\ref{bc}) describing two boundaries are also converted to $\xi=\xi_{s}\equiv\xi_0/c_s$, with $s=1,2$ ($\xi_1 < \xi_2$).

Obviously, here, we face a conformally trivial situation; a conformally invariant field, verifying Eq. (\ref{Eq}), on the conformally flat background (see (\ref{conformally})).\footnote{Respecting the fact that Dirichlet boundary condition is conformally invariant, a Dirichlet scalar in the curved bulk is corresponding to a Dirichlet scalar in a flat space-time.} Therefore, in the absence of the manifestation of the gravitational background\footnote{Note that for our case, a conformally coupled field on a conformally flat space-time, the free field vacuum expectation value of the energy-momentum tensor (the manifestation of the gravitational background) is completely determined by the trace anomaly \cite{Birrell}.} in the context of the Krein method, which actually corresponds to the situation without boundaries, the corresponding mean values of the energy-momentum tensor in this conformally trivial situation, can be evaluated by considering the standard transformation formula for conformally related problems from the related calculations in Rindler space as follows \cite{Birrell}
\begin{eqnarray}\label{Tmunu}\langle T_i^j[g'_{lm},\varphi^{{FRW}}] \rangle = [\xi/a(\eta)]^{D+1} \langle T_i^j[{\bar{g}}_{lm},\varphi^{{Rindler}}]\rangle.\hspace{0.5cm}\end{eqnarray}

In this paper, Rindler coordinates $(\tau,\xi,\textbf{x})$, related to Minkowskian ones $(t,x_1,\textbf{x})$, are considered
\begin{eqnarray}\label{rind=minko} t=\xi\sinh\tau, \;\; x_1 = \xi\cosh\tau,\end{eqnarray}
in which the set of coordinates parallel to the plates is characterized by $\textbf{x} = (x_2, ... ,x_D)$. In these coordinates, a wordline introduced by $\xi,\textbf{x} = constant$ determines an observer with constant proper acceleration $\xi^{-1}$. The Minkowski line element restricted to the Rindler wedge is
\begin{eqnarray}\label{metricrindler}ds^2 = \xi^2 d\tau^2 - d\xi^2 -d\textbf{x}^2.\end{eqnarray}

To make our notation more obvious, we first calculate the free field vacuum expectation value (VEV) of the energy-momentum tensor through the Krein method. The problem symmetry\footnote{The metric and the boundary conditions (which will be imposed on the field later) are static and translational-invariant in the hyperplane parallel to the plates.} allows us to write the corresponding part of the eigenfunctions in the standard plane wave structure as follows
\begin{eqnarray}\label{solution} &\bar\phi_\alpha(x) = C\bar\phi(\xi) e^{(i\textbf{kx}-i\omega\tau)},&\nn\\
&\alpha=(\textbf{k},\omega),\;\; \textbf{k}=(k_2, ... , k_d).\end{eqnarray}
Substituting this into Eq. (\ref{Eq}), the equation for $\bar\phi(\xi)$ on the background of the metric (\ref{metricrindler}) will be obtained as
\begin{eqnarray}\label{Eqrindler} \xi\frac{d}{d\xi}(\xi\frac{d\bar\phi}{d\xi}) + (\omega^2 - k^2\xi^2)\bar\phi(\xi)=0,\;\; k=|\textbf{k}|.\end{eqnarray}
The Bessel modified functions $\bar\phi(\xi)= I_{i\omega}(k\xi)$ and $K_{i\omega}(k\xi)$ (with the imaginary order) are actually the corresponding linearly independent solutions.

With regard to the metric (\ref{metricrindler}), the standard Klein-Gordon orthonormality condition for the eigenfunctions is defined by
\begin{eqnarray}\label{innerrindler} (\bar\phi_\alpha,\bar{\phi}'_\alpha)= -i \int d\textbf{x} \int \frac{d\xi}{\xi}\bar\phi_\alpha {\mathop{\partial_\tau} \limits^{\leftrightarrow}} {\bar\phi}^\ast_{\alpha'}.\end{eqnarray}
This orthonormality condition actually determines the coefficient $C$ in formula (\ref{solution}), so that, for the right Rindler wedge and in the absence of the boundaries, the considered eigenfunctions (\ref{solution}) take the following form (see, for instance, \cite{Fulling2850,Davies609,Unruh870,Candelas2101})
\begin{eqnarray}\label{modfree} \bar\phi_\alpha = \frac{\sqrt{\sinh\pi\omega}}{(2\pi)^{(D-1)/2}\pi} K_{i\omega} (k\xi) e^{(i\textbf{kx}-i\omega\tau)}.\end{eqnarray}
A complete set of mode solutions $\{ \bar\phi_\alpha,\bar\phi_\alpha^* \}$ of Eq. (\ref{Eqrindler}), which are orthonormal in the product (\ref{innerrindler}), are given by
\begin{eqnarray} &(\bar\phi_\alpha,\bar\phi_{\alpha'})=\delta _{\alpha\alpha'}, \;\; (\bar\phi_\alpha^*,\bar\phi_{\alpha'}^*)= -\delta _{\alpha\alpha'},& \nn\\
&(\bar\phi_\alpha,\bar\phi_{\alpha'}^*)= 0.& \label{eq:orthorindler} \end{eqnarray}

As already discussed, the field operator in the KGB quantization is
\begin{eqnarray}\label{fieldrindler} &\varphi^{{Rindler}} \equiv\bar\varphi = \frac{1}{\sqrt{2}}({\bar\varphi}_+ + {\bar\varphi}_-)&\nn\\
&= \int d^D\textbf{k} ({a}_\textbf{k} \bar\phi_\alpha + {a}^\dagger_\textbf{k} \bar\phi^*_\alpha) + \int d^D \textbf{k} ({b}^\dagger_\textbf{k} \bar\phi_\alpha + {b}_\textbf{k} \bar\phi^*_\alpha),\;\;&\end{eqnarray}
$[{a}_\textbf{k}, {a}^\dagger_{\textbf{k}'}] = \delta (\textbf{k} - \textbf{k}') ,\;\; [{b}_\textbf{k}, {b}^\dagger_{\textbf{k}'}] = -\delta (\textbf{k} - \textbf{k}')$, the other commutation relations are zero. The Fock vacuum state $|0\rangle$, for the right Rindler wedge, is defined by ${a}_\textbf{k} |{0}\rangle=0,\; {b}_\textbf{k} |{0}\rangle=0$.

With regard to the problem symmetry, the expression for the energy-momentum tensor of the considered scalar field can be written as \cite{Birrell}
$$T_{ij} = \nabla_i\bar\phi \nabla_j\bar\phi + [(\frac{D-1}{4D} - \frac{1}{4}) \bar{g}_{ij} \nabla_l\nabla^l - \frac{D-1}{4D}\nabla_i\nabla_j]\bar\phi^2.$$
The VEV of the energy-momentum tensor then will be
\begin{eqnarray}\label{<t>}
\langle{0}|T_{ij}(x)|{0}\rangle = \lim_{x\rightarrow x'} \nabla_i\nabla'_j G(x,x')\hspace{2.5cm}\nn\\
+ [(\frac{D-1}{4D} - \frac{1}{4}) \bar{g}_{ij} \nabla_l\nabla^l - \frac{D-1}{4D}\nabla_i\nabla_j]\langle{0}|\bar\varphi^2(x)|{0}\rangle.
\end{eqnarray}
The VEVs of the field square and the energy-momentum tensor can be evaluated in the sense of the Wightman function $G(x,x')$. Respecting the Krein quantum field (\ref{fieldrindler}), the corresponding Wightman function is
\begin{eqnarray}\label{0++-}G(x,x') &=& \langle{0}|\bar\varphi(x)\bar\varphi(x')|{0}\rangle \nn\\
&=&\langle{0}|\bar\varphi_+(x)\bar\varphi_+(x')|{0}\rangle + \langle{0}|\bar\varphi_-(x)\bar\varphi_-(x')|{0}\rangle,\;\;\;\;\;\;\end{eqnarray}
in which \cite{takookIJTP}
\begin{eqnarray}\label{0+=--*} \langle{0}|\bar\varphi_-(x)\bar\varphi_-(x')|{0}\rangle = -\Big(\langle{0}|\bar\varphi_+(x)\bar\varphi_+(x')|{0}\rangle\Big)^\ast,\end{eqnarray}
where
\begin{eqnarray}\label{0+}
\langle{0}|\bar\varphi_+(x)\bar\varphi_+(x')|{0}\rangle =  \frac{1}{\pi^2} \int\frac{d\textbf{k}}{(2\pi)^{D-1}} e^{i\textbf{k}(\textbf{x} - \textbf{x}')}\nn\\
\times \int_{0}^{\infty} d\omega  \sinh(\pi\omega) K_{i\omega}(k\xi)K_{i\omega}(k\xi')e^{-i\omega(\tau-\tau')}.\end{eqnarray}
Here, one can easily see that by substituting (\ref{0+=--*}) and (\ref{0+}) into formula (\ref{<t>}), actually by calculating the VEV of the energy-momentum tensor through the Krein method, due to the cancellations between the positive and negative norm parts of the field, the VEV of the energy-momentum tensor (the situation without boundaries) vanishes.

Now, let us study the effect of two boundaries given in (\ref{bc}) on the mean values of the energy-momentum tensor. In order to evaluate the Casimir energy-momentum tensor for Rindler space-time, it is convenient to divide the right Rindler wedge into three regions (it is assumed that the boundaries are situated in the right Rindler wedge, $x_1>|t|$): the regions outside the boundaries determined by $\xi<\xi_1$ and $\xi>\xi_2$ for which the mean values are the same as in the geometry with a single boundary, and the region between the boundaries $\xi_1<\xi<\xi_2$. In what follows, we will calculate the mean values of the energy-momentum tensor in these three regions individually.

\subsection{The region $\xi>\xi_2$}
By imposing Dirichlet boundary condition, $\bar\phi \Big|_{\xi=\xi_2} = 0$, with regard to the unitarity condition \cite{Garidi},\footnote{It has been shown that in the Minkowskian limit, especially when a theory with interaction is taken into account, through this quantization scheme and with respect to the unitarity condition, the results of the usual QFT can be recovered \cite{Garidi}.} the influence of applying physical boundary condition on the field operator is only upon the physical states. Accordingly, when physical boundary conditions are present, the field operator will be
\begin{eqnarray}\label{fieldrindlerbound}\bar\varphi =  \sum_n ({a}'_{\textbf{k}_n} \bar\phi'_\alpha + {a}'^\dagger_{\textbf{k}_n} \bar\phi'^*_\alpha)+ \int d^D \textbf{k} ({b}^\dagger_\textbf{k} \bar\phi_\alpha + {b}_\textbf{k} \bar\phi^*_\alpha),\;\;\end{eqnarray}
in which $ \bar\phi'_\alpha(x) = C' K_{i\omega_n} (k\xi) e^{(i\textbf{kx}-i\omega_n\tau)}$. Note that the normalization condition (\ref{innerrindler}), while the $\xi$-integration goes over the region $(\xi_2,\infty)$, specifies the coefficient $C'$ \cite{Saharian}
\begin{eqnarray}\label{C'} C'^2 = \frac{1}{(2\pi)^{D-1}} \frac{I_{i\omega_n}(k\xi_2)}{\frac{\partial}{\partial\omega} K_{i\omega}(k\xi_2)\Big|_{\omega=\omega_n}},\end{eqnarray}
$\omega_n$ are indeed the possible values for $\omega$ verifying $K_{i\omega}(k\xi_2)=0$.

Considering the above construction, the physical part of the Wightman function would be
\begin{eqnarray}\label{+}
\langle\bar\varphi'_+(x)\bar\varphi'_+(x')\rangle_{\xi>\xi_2} = \int\frac{d\textbf{k}}{(2\pi)^{D-1}} e^{i\textbf{k}(\textbf{x} - \textbf{x}')}\hspace{1.5cm}\nn\\
\times\sum_{n=1}^{\infty} \frac{I_{i\omega}(k\xi_2)}{\frac{\partial}{\partial\omega} K_{i\omega}(k\xi_2)} K_{i\omega}(k\xi)K_{i\omega}(k\xi')e^{-i\omega(\tau-\tau')}\Big|_{\omega=\omega_n},\;\;\end{eqnarray}
whereas, respecting the unitarity condition, the non-physical part remains unaffected and still can be identified by (\ref{0+=--*}).

By making use of the generalized Abel-Plana formula given in Ref. \cite{Saharian}, for a function $W(z)$ which is assumed to be an analytic function in the right half-plane ($\omega_{n_l} <l< \omega_{n_{l+1}}$)
$$\lim_{l\rightarrow\infty}\Big\{ \sum_{n=1}^{n_l} \frac{I_{i\omega_n}(\eta)W(\omega_n)}{\partial K_{i\omega}(\eta)/\partial\omega\Big|_{\omega=\omega_n}} - \frac{1}{\pi^2}\int_0^l \sinh(\pi z) W(z)dz \Big\}$$$$ = - \frac{1}{2\pi} \int_0^\infty \frac{I_z(\eta)}{K_z(\eta)} [ W(xe^{i\pi/2}) + W(ze^{-i\pi/2}) ]dz,$$
one can interestingly rewrite physical part of the Wightman function, (\ref{+}), in the following form
\begin{eqnarray}\label{++}
\langle\bar\varphi'_+(x)\bar\varphi'_+(x')\rangle_{\xi>\xi_2}=\hspace{5.2cm} \nn\\
\langle{0}|\bar\varphi_+(x)\bar\varphi_+(x')|{0}\rangle - \frac{1}{\pi} \int \frac{d\textbf{k}}{(2\pi)^{D-1}} e^{i\textbf{k}(\textbf{x}-\textbf{x}')}\hspace{0.5cm}\nn\\
\times\int_0^\infty d\omega\frac{I_\omega(k\xi_2)}{K_\omega(k\xi_2)} K_\omega(k\xi)K_\omega(k\xi') \cosh[\omega(\tau-\tau')]. \;\;\end{eqnarray}
Note that, all divergences in the coincidence limit are contained in the first term corresponding to the situation without boundaries (given in (\ref{0+})).

Now, by substituting the Wightman function obtained for the region outside the plates $\xi>\xi_2$ (including the physical part, see Eq. (\ref{++}), and the non-physical part, see Eq. (\ref{0+=--*})), into Eq. (\ref{<t>}), the Casimir energy-momentum tensor can be obtained as follows
\begin{widetext}
\begin{eqnarray}\label{<t>''}
\langle T_{i}^j\rangle_{\xi>\xi_2} = \frac{-\delta_i^j}{2^{D-2}\pi^{(D+1)/2} \Gamma(\frac{D-1}{2})} \int_0^\infty dk k^D \int_0^\infty d\omega \frac{I_\omega(k\xi_2)}{K_\omega(k\xi_2)} F^{(i)}[{K_\omega(k\xi)}],
\end{eqnarray}
in which
\begin{equation} \label{<t>''0} F^{(0)}[{K_\omega(k\xi)}] = \frac{1}{2D} \Big[ \Big(\frac{d{K_\omega(k\xi)}}{k d\xi} \Big)^2 + \Big( 1+\frac{\omega^2}{k^2\xi^2} \Big){K^2_\omega(k\xi)}\Big] + \frac{D-1}{4Dk\xi}\frac{d}{k d\xi}{K^2_\omega(k\xi)} - \frac{\omega^2}{k^2\xi^2} {K^2_\omega(k\xi)},
\end{equation}
\begin{equation} \label{<t>''1} F^{(1)}[{K_\omega(k\xi)}] = - \frac{1}{2} \Big(\frac{d{K_\omega(k\xi)}}{k d\xi} \Big)^2 - \frac{D-1}{4Dk\xi}\frac{d}{k d\xi}{K^2_\omega(k\xi)} + \frac{1}{2} \Big(1+\frac{\omega^2}{k^2\xi^2}\Big){K^2_\omega(k\xi)},
\end{equation}
\begin{eqnarray} \label{<t>''i} F^{(i)}[{K_\omega(k\xi)}] = \frac{1}{2D} \Big[ \Big(\frac{d{K_\omega(k\xi)}}{k d\xi} \Big)^2 + \Big( 1+\frac{\omega^2}{k^2\xi^2}\Big){K^2_\omega(k\xi)}\Big] - \frac{{K^2_\omega(k\xi)}}{D-1}, \;\;\; i=2, 3, ..., D,
\end{eqnarray}
\end{widetext}
where, indices $0$ and $1$ correspond to the coordinates $\tau$ and $\xi$, respectively. It can be easily seen that for our case, a conformally coupled massless scalar field, the energy-momentum tensor is traceless.

Here, it must be emphasized that, due to the presence of non-physical states in the context of the Krein quantization scheme, the divergent part of the physical part of the Casimir energy-momentum tensor (related to the first term on the right side of Eq. (\ref{++})) is automatically eliminated and the final result is finite. This procedure is similar to that used by Candelas and Deutsch \cite{PCandelas354}.

\subsection{The region $\xi<\xi_1$}
Calculating the Casimir energy-momentum tensor in this region could be easily performed by considering the related one in the region $\xi>\xi_2$, i.e. $\langle T_{i}^j\rangle_{\xi>\xi_2}$ (see (\ref{<t>''})). Actually, this situation is the same as for the interior and exterior regions in the case of cylindrical and spherical surfaces on background of the Minkowski space-time. This similarity is due to the fact that the uniformly accelerated observers create worldlines in Minkowski space-time that would correspond to circles in Euclidean space. This correspondence extends also to the field equation and its solutions, and modes (\ref{solution}) are the Minkowski space-time analogous corresponding to cylinder harmonics. On this basis, mean values of the energy-momentum tensor in the region $\xi<\xi_1$, can be easily obtained by the following replacements \cite{Saharian125007,Romeo105019}
\begin{eqnarray}\label{replacements} \langle T_{i}^j\rangle_{\xi<\xi_1} = \Big\{\langle T_{i}^j\rangle_{\xi>\xi_2};\; \xi_2 \rightarrow \xi_1, \; I_\omega\rightleftarrows K_\omega\Big\}. \end{eqnarray}

\subsection{The region $\xi_1<\xi<\xi_2$}
In the region between the plates the linearly independent solutions to equation (\ref{Eqrindler}) verifying Dirichlet boundary condition on the plate $\xi=\xi_2$ are
\begin{eqnarray}\label{D2} D_{i\omega}(k\xi,k\xi_2) =I_{i\omega}(k\xi_2)K_{i\omega}(k\xi) - K_{i\omega}(k\xi_2)I_{i\omega}(k\xi).\;\;\end{eqnarray}
Note that, $D_{i\omega}(k\xi,k\xi_2) = D_{-i\omega}(k\xi,k\xi_2)$. Considering Dirichlet boundary condition on the plate $\xi=\xi_1$, the allowed values for $\omega$; $\omega_n= \omega_n (k\xi_1,k\xi_2)>0$, $n=1,2,..$, can be obtained by
\begin{eqnarray}\label{D1} D_{i\omega}(k\xi_1,k\xi_2) =0.\end{eqnarray}
For a fixed value of $k$, this equation has an infinite set of real solutions \cite{Saharian2353}. Here, it is assumed that they are arranged in the ascending order $\omega_n < \omega_{n+1}$.

In the region between the plates, and with respect to the standard Klein-Gordon orthonormality condition (\ref{innerrindler}), the solutions induced by boundaries are \cite{Saharian2353}
\begin{eqnarray}\label{solution12} \bar\phi''_\alpha = C''D_{i\omega}(k\xi,k\xi_2) e^{i(\textbf{kx}-\omega_n\tau)},\end{eqnarray}
in which
\begin{eqnarray}\label{C'''} C''^2= \frac{(2\pi)^{1-D} I_{i\omega}(k\xi_1)}{I_{i\omega}(k\xi_2)\frac{\partial}{\partial\omega}D_{i\omega}(k\xi_1,k\xi_2)}\Big|_{\omega=\omega_n}.\end{eqnarray}

Respecting the Krein quantum field, therefore, the physical part of the Wightman function for the region between the plates is as follows
\begin{eqnarray}\label{+''} \langle\bar\varphi''_+(x)\bar\varphi''_+(x')\rangle_{\xi_1<\xi<\xi_2} = \hspace{4cm} \nn\\
\int\frac{d\textbf{k}e^{i\textbf{k}(\textbf{x}-\textbf{x}')}}{(2\pi)^{D-1}}
\sum_{n=1}^{\infty}\frac{I_{i\omega}(k\xi_1)e^{-i\omega(\tau-\tau')}}{I_{i\omega}(k\xi_2)\frac{\partial}{\partial\omega}D_{i\omega}(k\xi_1,k\xi_2)}\nn\\
\times D_{i\omega}(k\xi,k\xi_2)D_{i\omega}(k\xi',k\xi_2)\Big|_{\omega=\omega_n}.\;\;
\end{eqnarray}
Again, respecting the unitarity condition, the non-physical part do not interact with the physical states or real physical world, hence it cannot be affected by the physical boundary conditions. Therefore, the non-physical part of the Wightman function still can be considered as (\ref{0+=--*}).

The physical part of the Wightman function, respecting the procedure given in Ref. \cite{Saharian2353}, and using the generalized Abel-Plana formula, can be written as
\begin{eqnarray}\label{+''new} \langle\bar\varphi''_+(x)\bar\varphi''_+(x')\rangle_{\xi_1<\xi<\xi_2} = \langle\bar\varphi'_+(x)\bar\varphi'_+(x')\rangle_{\xi<\xi_2}\hspace{1cm}\nn\\
- \int\frac{d\textbf{k}e^{i\textbf{k}(\textbf{x}-\textbf{x}')}}{\pi(2\pi)^{D-1}} \int_0^\infty d\omega \frac{I_{\omega}(k\xi_1)}{I_{\omega}(k\xi_2)D_{\omega}(k\xi_1,k\xi_2)}\hspace{0.7cm}\nn\\
\times D_\omega(k\xi,k\xi_2) D_\omega(k\xi',k\xi_2) \cosh[\omega(\tau-\tau')], \;\;\;
\end{eqnarray}
in which, the first term on the right-hand side stands for the physical part of the Wightman function in the region ${\xi<\xi_2}$ for a single plate at ${\xi=\xi_2}$. Respecting the statements given in subsection B, it can be easily obtained by substituting $I_\omega\rightleftarrows K_\omega$ in (\ref{++}).

The physical part of the Wightman function (\ref{+''}) can be also presented in the form \cite{Saharian2353}
\begin{eqnarray}\label{+''new'} \langle\bar\varphi''_+(x)\bar\varphi''_+(x')\rangle_{\xi_1<\xi<\xi_2} = \langle\bar\varphi'_+(x)\bar\varphi'_+(x')\rangle_{\xi_1<\xi}\hspace{1.5cm}\nn\\
- \int\frac{d\textbf{k}e^{i\textbf{k}(\textbf{x}-\textbf{x}')}}{\pi(2\pi)^{D-1}} \int_0^\infty d\omega \frac{K_{\omega}(k\xi_2)}{K_{\omega}(k\xi_1)D_{\omega}(k\xi_1,k\xi_2)}\hspace{0.7cm}\nn\\
\times D_\omega(k\xi,k\xi_1) D_\omega(k\xi',k\xi_1) \cosh[\omega(\tau-\tau')], \;\;\;
\end{eqnarray}
here, the first term refers to the physical part of the Wightman function in the region ${\xi_1<\xi}$ for a single plate at ${\xi=\xi_1}$. It can be easily obtained by substituting $\xi_2\rightarrow\xi_1$ in (\ref{++}).

Now, by substituting the calculated Wightman function for the region between the plates (including the physical part, see Eqs. (\ref{+''new}) and (\ref{+''new'}), and the non-physical part, see Eq. (\ref{0+=--*})), into the (\ref{<t>}), we obtain two equivalent representations for the $\langle T_i^j\rangle_{{\xi_1<\xi<\xi_2}}$,
\begin{eqnarray}\label{<>} \langle T_i^j\rangle_{{\xi_1<\xi<\xi_2}} = \langle T_i^j\rangle_{{\xi<\xi_2}} - \frac{\delta_i^j}{2^{D-2}\pi^{\frac{D+1}{2}}\Gamma(\frac{D-1}{2})} \int dk k^D \hspace{0.5cm}\nn\\
\times \int_0^\infty d\omega\frac{I_{\omega}(k\xi_1)}{I_{\omega}(k\xi_2)D_{\omega}(k\xi_1,k\xi_2)}F^{(i)}[D_\omega(k\xi,k\xi_2)],\;\;\;\end{eqnarray}
and
\begin{eqnarray}\label{<>'} \langle T_i^j\rangle_{{\xi_1<\xi<\xi_2}} = \langle T_i^j\rangle_{{\xi_1<\xi}} - \frac{\delta_i^j}{2^{D-2}\pi^{\frac{D+1}{2}}\Gamma(\frac{D-1}{2})} \int dk k^D \hspace{0.5cm}\nn\\
\times \int_0^\infty d\omega\frac{K_{\omega}(k\xi_2)}{K_{\omega}(k\xi_1)D_{\omega}(k\xi_1,k\xi_2)}F^{(i)}[D_\omega(k\xi,k\xi_1)],\;\;\;\end{eqnarray}
in which, with respect to (\ref{<t>''}), we have
\begin{eqnarray}\label{1}
\langle T_i^j\rangle_{{\xi_1<\xi}} = \Big\{ \langle T_i^j\rangle_{{\xi_2<\xi}};\; \xi_2 \rightarrow\xi_1\Big\},
\end{eqnarray}
\begin{eqnarray}\label{2}
\langle T_i^j\rangle_{{\xi<\xi_2}} = \Big\{ \langle T_i^j\rangle_{{\xi_2<\xi}};\; I_\omega\rightleftarrows K_\omega\Big\},
\end{eqnarray}
and according to (\ref{<t>''0})-(\ref{<t>''i}),
\begin{eqnarray}\label{333} F^{(i)}[D_\omega(k\xi,k\xi_{s=1,2})]\hspace{4.5cm}\nn\\ = \Big\{ F^{(i)}[{K_\omega(k\xi)}]; K_\omega(k\xi) \rightarrow D_\omega(k\xi,k\xi_{s=1,2}) \Big\}.\end{eqnarray}

Note that, the form (\ref{<>}) (respectively (\ref{<>'})) is convenient to study the expectation values near the plate  ${\xi=\xi_2}$ (respectively ${\xi=\xi_1}$). As a matter of fact, the first term on the right of (\ref{<>}) (resp. (\ref{<>'})) is finite on the plate ${\xi=\xi_2}$ (resp. ${\xi=\xi_1}$) and diverges on the plate at ${\xi=\xi_1}$ (resp. ${\xi=\xi_2}$).

\subsection{Coming back to the FRW space-time;\\evaluating the Casimir energy-momentum tensor}
Thus far, the Casimir energy-momentum tensor in Rindler space-time induced by Dirichlet boundary condition on two boundaries described by $\xi=\xi_1, \xi_2$ has been calculated. The corresponding Casimir energy-momentum tensor for the FRW space-time, in $x'$ coordinates, can be easily obtained with regard to the conformal relation between these spaces, see Eq. (\ref{Tmunu}). Moreover, by using the inverse coordinate transformation $x'^i \rightarrow x^i$, one can calculate the Casimir energy-momentum tensor in FRW hyperspherical coordinates (see Eq. (\ref{metric})). Technically, for a rank-2 tensor $S_{ij}$, which is diagonal in coordinates $x'^i = (\eta,x')$, the transformation to coordinates $x^i = (\eta,r,\theta_1, ... , \theta_{D-1})$ has the following form (no summation over $j$)
\begin{eqnarray}\label{inv-ct-coeff}
S_0^0 & = & {S'}_0^0, \nn\\
S_1^1 & = & {S'}_1^1 + \Omega^2 \sin^2 \theta ({S'}_2^2 - {S'}_1^1), \nn\\
S_1^2 & = & \Omega^2 \sin \theta \frac{\cos \theta - r\gamma}{r}({S'}_2^2 - {S'}_1^1), \nn\\
S_2^2 & = & {S'}_2^2 + \Omega^2 \sin^2 \theta ({S'}_1^1 - {S'}_2^2), \nn\\
S_j^j & = & {S'}_2^2, \;\; j = 3, ..., D.
\end{eqnarray}

After applying this coordinate transformation, the boundary-induced energy-momentum tensor in coordinates $x^i$ will be obtained as follows (no summation over $j$)
\begin{widetext}
\begin{eqnarray}\label{entensor}
\langle T_j^j \rangle^{FRW} & = & [\xi/a(\eta)]^{D+1} \langle T_j^j \rangle^{Rindler}, \;\; j = 3, ..., D, \nn\\
\langle T_j^j \rangle^{FRW} & = & [\xi/a(\eta)]^{D+1} \Big[ \langle T_j^j \rangle^{Rindler} + (-1)^j \Omega^2 \sin^2 \theta \left( \langle T_1^1 \rangle^{Rindler} - \langle T_2^2 \rangle^{Rindler} \right) \Big], \;\; j = 1,2, \nn\\
\langle T_1^2 \rangle^{FRW} & = & [\xi/a(\eta)]^{D+1} \Omega^2 \sin \theta \frac{\cos \theta - r\gamma}{r} \left( \langle T_2^2 \rangle^{Rindler} - \langle T_1^1 \rangle^{Rindler} \right),
\end{eqnarray}
\end{widetext}
in which $\langle T_i^j \rangle^{Rindler}$ is the corresponding energy-momentum tensor for each region introduced in the above subsections (see Eqs. (\ref{<t>''}), (\ref{replacements}) and (\ref{<>})). Note that, due to the fact that the spatial part of the boundary-induced energy-momentum tensor in coordinates $x'^i$ is not isotropic, the obtained result for the corresponding part in coordinates $x^i$ is not diagonal.

Now, possessing the VEVs of the energy-momentum tensor, we can study the Casimir forces acting on the boundaries. Before that, let us define the normal to the boundary. In the region between the boundaries, the corresponding normal has the components
\begin{equation} \label{normal}
n_s^i = \delta_s[c_sra(\eta)]^{-1}(0, \gamma^2(\gamma - c_s), -\sin\theta, 0, ..., 0),
\end{equation}
where $\delta_1 = 1$ and $\delta_2 = -1$.

The j-th component of the Casimir force acting per unit surface of the boundary at $\xi = \xi_s$ is specified by the expression $\Big( \langle T_i^j \rangle^{FRW} \Big) n_s^i |_{\xi = \xi_s}$. Utilizing (\ref{normal}) and (\ref{entensor}), then the force can be expressed in the following form
\begin{eqnarray} \label{1212}
\Big(\langle T_i^j \rangle^{FRW} \Big) n_s^i |_{\xi = \xi_s}\hspace{5cm}\nn\\
\hspace{1cm}= n_s^j[\xi_s/a(\eta)]^{D+1} \Big(\langle T_1^1 \rangle^{Rindler}\Big) |_{\xi = \xi_s}.\;\;
\end{eqnarray}
Note that the quantity
$$\Big( P^{Rindler} \Big)^{(s)} = -\Big( \langle T_1^1 \rangle^{Rindler} \Big)|_{\xi = \xi_s},$$
specifies the pressure on the plate at $\xi = \xi_s$ in the corresponding Rindler problem. It can be presented as a sum of two terms; the pressure for a single plate at $\xi = \xi_s$ when the second plate is absent (the first term on the right) and the pressure induced by the presence of the second plate (the second term on the right)
\begin{eqnarray} \label{1313}
\Big( P^{Rindler} \Big)^{(s)} = \Big(p_1^{Rindler}\Big)^{(s)} + \Big( P^{Rindler}_{(int)} \Big)^{(s)},
\end{eqnarray}
with $s=1,2$. The first term is divergent due to the well-known surface divergences in the subtracted VEV's (see \cite{Milton}). Contrary to the previous one, the interaction parts are finite for all nonzero distances between the boundaries.

For the plate at ${\xi=\xi_2}$ the interaction term is due to the second term on the right of Eq. (\ref{<>}). Employing the Wronskian relation for the modified Bessel functions,\footnote{$I_{\nu}(x)K'_{\nu}(x) - K_{\nu}(x)I'_{\nu}(x) = -\frac{1}{x}.$} when ${\xi=\xi_2}$, this term is converted to
\begin{eqnarray}\label{<>""} \Big(p^{Rindler}_{(int)}\Big)^{(2)} = - \frac{\delta_i^j}{2^{D-1}\xi^2_2\pi^{\frac{D+1}{2}}\Gamma(\frac{D-1}{2})} \int dk k^{D-2} \hspace{0.5cm}\nn\\
\times \int_0^\infty d\omega\frac{I_{\omega}(k\xi_1)}{I_{\omega}(k\xi_2)D_{\omega}(k\xi_1,k\xi_2)}.\;\;\;\end{eqnarray}
Pursuing the same path for the interaction term on the plate at ${\xi=\xi_1}$, the second term on the right of Eq. (\ref{<>'}), we have
\begin{eqnarray}\label{<>""''} \Big(p^{Rindler}_{(int)}\Big)^{(1)} = - \frac{\delta_i^j}{2^{D-1}\xi^2_1\pi^{\frac{D+1}{2}}\Gamma(\frac{D-1}{2})} \int dk k^{D-2} \hspace{0.5cm}\nn\\
\times \int_0^\infty d\omega\frac{K_{\omega}(k\xi_2)}{K_{\omega}(k\xi_1)D_{\omega}(k\xi_1,k\xi_2)}.\;\;\;\end{eqnarray}
Now, the corresponding pressure in FRW bulk can be easily obtained by using the above expressions, (\ref{<>""}) and (\ref{<>""''}),
\begin{eqnarray}
\Big(P^{FRW}_{(int)} \Big)^{(s)} = [\xi_s/a(\eta)]^{D+1}\Big(p^{Rindler}_{(int)}\Big)^{(s)},
\end{eqnarray}
Note that the function $D_{\omega}(k\xi_1,k\xi_2)$ is positive for $\xi_1<\xi_2$, therefore, interaction forces per unit surface (\ref{<>""}) and (\ref{<>""''}) are always attractive. They are finite for all $\xi_1<\xi_2$, and do not depend on the curvature coupling parameter; $\frac{D-1}{4D}$ (see Eq. (\ref{Eq})).

\section{Conclusion}
In summary, considering a brane-world setup when two $D$-dimensional branes are embedded in $D+1$-dimensional FRW space-time (with negative spatial curvature), we have calculated mean values of the energy-momentum tensor for a quantized bulk scalar field coupled conformally to the curvature in FRW background. The branes have been chosen to be the conformal images of two infinite parallel plates moving with constant proper acceleration in Rindler space-time, while admitting Dirichlet boundary condition on their surfaces. Technically, the Krein-Gupta-Bleuler construction as a respectful quantization scheme in (asymptotically) dS space-times has been considered to perform the calculations.

In this context, having mean values of the energy-momentum tensor, we have also evaluated the Casimir forces and effective pressure acting on the branes. These Casimir forces consist of two parts; self-action and interaction parts. The interaction forces are directed along the normal to the boundary and are independent of the point of the boundary and the curvature coupling parameter. In addition, they are attractive for all separations between the boundaries.

\section*{Acknowledgements}
The authors would like to thank the referee for instrumental comments.

\end{document}